\RequirePackage{fix-cm}
\documentclass[smallextended]{svjour3}       
\smartqed  
\usepackage{graphicx}
\begin{document}

\title{Spontaneously emitted X-rays: an experimental signature of the dynamical reduction models
}
\subtitle{}


\author{C.~Curceanu        \and
        S.~Bartalucci\and A.~Bassi\and M.~Bazzi \and S.~Bertolucci\and C.~Berucci\and A.~M.~Bragadireanu\and M.~Cargnelli\and A.~Clozza\and L.~De~Paolis\and S.~Di~Matteo\and S.~Donadi\and A.~D'Uffizi\and J-P.~Egger\and C.~Guaraldo\and M.~ Iliescu\and T.~ Ishiwatari\and M.~ Laubenstein\and J.~ Marton\and E.~ Milotti\and A.~Pichler \and D.~Pietreanu\and K.~Piscicchia\and T~.Ponta\and E.~Sbardella\and A.~Scordo\and H.~Shi\and D.L.~Sirghi\and F.~Sirghi\and L.~Sperandio\and O.~Vazquez~Doce\and J.~Zmeskal
}


\institute{C. Curceanu \at
INFN, Laboratori Nazionali di Frascati, CP 13, Via E. Fermi 40, I-00044, Frascati (Roma), Italy
              \\
              Tel.: +390694032321\\
              Fax: +390694032559\\
              \email{catalina.curceanu@lnf.infn.it}  \\
Museo Storico della Fisica e Centro Studi e Ricerche ``Enrico Fermi'', Roma, Italy\\
``Horia Hulubei''National Institute of Physics and Nuclear Engineering, Str. Atomistilor no. 407, P.O. Box MG-6,  Bucharest - Magurele, Romania
\and
           S. Bartalucci\at
INFN, Laboratori Nazionali di Frascati, CP 13, Via E. Fermi 40, I-00044, Frascati (Roma), Italy
\and
             A. Bassi\at
Dipartimento di Fisica, Universit\`{a} di Trieste and INFN-- Sezione di Trieste, Via Valerio, 2, I-34127 Trieste, Italy
\and
M. Bazzi\at
INFN, Laboratori Nazionali di Frascati, CP 13, Via E. Fermi 40, I-00044, Frascati (Roma), Italy
\and
S. Bertolucci\at
INFN, Laboratori Nazionali di Frascati, CP 13, Via E. Fermi 40, I-00044, Frascati (Roma), Italy
\and
C. Berucci\at
INFN, Laboratori Nazionali di Frascati, CP 13, Via E. Fermi 40, I-00044, Frascati (Roma), Italy\\
The Stefan Meyer Institute for Subatomic Physics,  Boltzmanngasse 3, A-1090 Vienna, Austria
\and
A. M. Bragadireanu\at
INFN, Laboratori Nazionali di Frascati, CP 13, Via E. Fermi 40, I-00044, Frascati (Roma), Italy\\
``Horia Hulubei''National Institute of Physics and Nuclear Engineering, Str. Atomistilor no. 407, P.O. Box MG-6,  Bucharest - Magurele, Romania
\and
M. Cargnelli\at
The Stefan Meyer Institute for Subatomic Physics,  Boltzmanngasse 3, A-1090 Vienna, Austria
\and
A. Clozza\at
INFN, Laboratori Nazionali di Frascati, CP 13, Via E. Fermi 40, I-00044, Frascati (Roma), Italy
\and
 L. De Paolis\at
 INFN, Laboratori Nazionali di Frascati, CP 13, Via E. Fermi 40, I-00044, Frascati (Roma), Italy
\and
S~.Di~Matteo\at
Institut de Physique UMR CNRS-UR1 6251, Universit\'e de Rennes1, F-35042 Rennes, France
\and
S. Donadi\at
Dipartimento di Fisica, Universit\`{a} di Trieste and INFN-- Sezione di Trieste, Via Valerio, 2, I-34127 Trieste, Italy
\and
A. D'Uffizi\at
INFN, Laboratori Nazionali di Frascati, CP 13, Via E. Fermi 40, I-00044, Frascati (Roma), Italy
\and
J-P. Egger\at
Institut de Physique, Universit\'e de Neuch\^atel, 1 rue A.-L. Breguet, CH-2000 Neuch\^atel, Switzerland
\and
C. Guaraldo\at
INFN, Laboratori Nazionali di Frascati, CP 13, Via E. Fermi 40, I-00044, Frascati (Roma), Italy
\and
M. Iliescu\at
INFN, Laboratori Nazionali di Frascati, CP 13, Via E. Fermi 40, I-00044, Frascati (Roma), Italy
\and
T. Ishiwatari\at
The Stefan Meyer Institute for Subatomic Physics,  Boltzmanngasse 3, A-1090 Vienna, Austria
\and
M. Laubenstein\at
INFN, Laboratori Nazionali del Gran Sasso, S.S. 17/bis, I-67010 Assergi (AQ), Italy
\and
J. Marton\at
The Stefan Meyer Institute for Subatomic Physics,  Boltzmanngasse 3, A-1090 Vienna, Austria
\and
E. Milotti\at
Dipartimento di Fisica, Universit\`{a} di Trieste and INFN-- Sezione di Trieste, Via Valerio, 2, I-34127 Trieste, Italy
\and
M. Pichler\at
The Stefan Meyer Institute for Subatomic Physics,  Boltzmanngasse 3, A-1090 Vienna, Austria
\and
D. Pietreanu\at
INFN, Laboratori Nazionali di Frascati, CP 13, Via E. Fermi 40, I-00044, Frascati (Roma), Italy\\
``Horia Hulubei''National Institute of Physics and Nuclear Engineering, Str. Atomistilor no. 407, P.O. Box MG-6,  Bucharest - Magurele, Romania
\and
K.~Piscicchia\at
INFN, Laboratori Nazionali di Frascati, CP 13, Via E. Fermi 40, I-00044, Frascati (Roma), Italy\\
Museo Storico della Fisica e Centro Studi e Ricerche ``Enrico Fermi'', Roma, Italy
\and
T. Ponta\at
``Horia Hulubei''National Institute of Physics and Nuclear Engineering, Str. Atomistilor no. 407, P.O. Box MG-6,  Bucharest - Magurele, Romania
\and
E. Sbardella\at
INFN, Laboratori Nazionali di Frascati, CP 13, Via E. Fermi 40, I-00044, Frascati (Roma), Italy
\and
A. Scordo\at
INFN, Laboratori Nazionali di Frascati, CP 13, Via E. Fermi 40, I-00044, Frascati (Roma), Italy
\and
H. Shi\at
INFN, Laboratori Nazionali di Frascati, CP 13, Via E. Fermi 40, I-00044, Frascati (Roma), Italy\\
The Stefan Meyer Institute for Subatomic Physics,  Boltzmanngasse 3, A-1090 Vienna, Austria
\and
D.L. Sirghi\at
INFN, Laboratori Nazionali di Frascati, CP 13, Via E. Fermi 40, I-00044, Frascati (Roma), Italy\\
``Horia Hulubei''National Institute of Physics and Nuclear Engineering, Str. Atomistilor no. 407, P.O. Box MG-6,  Bucharest - Magurele, Romania
\and
F. Sirghi\at
INFN, Laboratori Nazionali di Frascati, CP 13, Via E. Fermi 40, I-00044, Frascati (Roma), Italy\\
``Horia Hulubei''National Institute of Physics and Nuclear Engineering, Str. Atomistilor no. 407, P.O. Box MG-6,  Bucharest - Magurele, Romania
\and
 L.~Sperandio\at
INFN, Laboratori Nazionali di Frascati, CP 13, Via E. Fermi 40, I-00044, Frascati (Roma), Italy
\and
O. Vazquez Doce\at
Excellence Cluster Universe, Technische Universit\"at M\"unchen, Garching, Germany
\and
J. Zmeskal\at
The Stefan Meyer Institute for Subatomic Physics,  Boltzmanngasse 3, A-1090 Vienna, Austria
}

\date{Received: date / Accepted: date}

\maketitle

\begin{abstract}
We  present the idea of  searching for  X-rays as  a signature of the mechanism inducing the spontaneous collapse of the wave function. Such a signal is  predicted by the continuous spontaneous localization theories, which are solving the ``measurement problem'' by modifying the Schr\"odinger equation. We will show some encouraging preliminary results and discuss future plans and strategy.\\ 
PACS: 29.30.kv collapse models; 32.30.Rj X-ray measurements

\keywords{collapse models \and dynamical reduction models \and X-rays}
\end{abstract}

\section{Introduction}
\label{intro}

In spite of its tremendous success, Quantum Mechanics  still generates many discussions related to its possible limits.
In this paper we present an experimental test for one of the most debated items in quantum theory, namley the ``measurement problem'', within the spontaneous collapse of the wave function model.

The recent development of testable, mathematically complete and consistent models, the Dynamical Reduction Models, as a possible solution of the ``measurement problem'', strongly renewed the interest of the scientific community in the foundations of quantum mechanics. 
The first consistent and satisfying  Dynamical Reduction Model, known as Quantum Mechanics with Spontaneous Localisation (QMSL), was developed by Ghirardi, Rimini and Weber \cite{ghi}. According to this model, particles undergo spontaneous localisations around definite positions, following a Possion distribution characterised by a  mean frequency   $\lambda \sim 10^{-16}$ s$^{-1}$. The efforts of Ghirardi, Rimini, Weber and Pearle \cite{pear} brought to the development of the CSL model (Continuous Spontaneous Localisation), based on the introduction of new, non linear and stochastic terms in the Schr\"odinger equation, besides the standard ones. Such terms induce, for the state vector, a diffusion process, which is responsible for the wave packet reduction.  
The Dynamical Reduction Models posses the characteristic of being experimentally testable by measuring the predicted deviations with respect to standard quantum mechanics. The conventional approach to test the collapse models is to generate spatial superpositions of mesoscopic systems and examine the loss of interference, while environmental noises are under control. Naturally oscillating systems, such as neutrinos, neutral mesons and chiral molecules, create quantum superpositions and thus represent, as well, a natural case-study for testing the quantum linearity. However, the collapse models can not be tested with neutrinos and the effect, stronger for neutral mesons, is still beyond experimental reach, while chiral molecules can offer valid candidates for testing collapse models \cite{bahrami}. The most promising testing ground is offered by the spontaneous emission of radiation due to the particle interaction with the stochastic field which causes  an enhancement of the energy expectation value. This implies, for a free charged particle, the emission of electromagnetic radiation. Such a spontaneous radiation is  not present in the standard quantum mechanics. A measurement of the emitted radiation rate thus makes it possible to obtain valuable information on the $\lambda$ parameter of the collapse models.
Q. Fu \cite{fu} obtained, for the first time, an upper limit for $\lambda$ based on this effect, using  the radiation measured by a system of Germanium detectors\cite{miley}. 

In this paper we present a more refined analysis of the X-ray emission spectrum measured by the IGEX collaboration \cite{igex1} and improve the previous limit on $\lambda$  by a factor about 4. 

 We will conclude the paper by presenting some ideas to perform in the near future a  dedicated measurement and to gain more than one order of magnitude in the limit of the  $\lambda$-parameter.

 \section{The wave function localisation process in the spontaneous collapse models}
\label{wave}

Part of the interest of the scientific community for the Dynamical Reduction Models as a possible solution to the ``measurement problem'' stands in their  characteristic to be experimentally testable. The stochastic field  which is responsible for the reduction of the wave packet, as shown in  \cite{fu}, also causes an enhancement of the energy expectation value, which, for the charged particles, means an emission of electromagnetic radiation (called spontaneous radiation) not present in the standard quantum mechanics. The rate of this emitted radiation depends on the parameter $\lambda$ characterizing the collapse models. Consequently,  an upper limit for the mean collapse frequency   $\lambda$ can be established by the measurement of the emitted radiation rate.

The rate of the radiation spontaneously emitted by free electrons, as a consequence of their interaction with the stochastic field, was calculated in the framework of the non-relativistic CSL model, assuming that $\lambda$ is an universal parameter, not depending on the particle mass, and it is given by \cite{fu}:

\begin{equation}\label{furate}
\frac{d\Gamma (E)}{dE} = \frac{e^2 \lambda}{4\pi^2 a^2 m^2 E}
\end{equation}
where $m$ represents the electron mass, $E$ is the energy of the emitted photon, $\lambda$ and $a$ are, respectively, the reduction rate parameter and the correlation length of the reduction model. The $a$ parameter value is taken to be $a = 10^{-7} m$. Using the measured radiation rate obtained for an isolated Germanium detector \cite{miley} Fu extracted the following upper limit for $\lambda$:

\begin{equation}\label{fulimit}
\lambda < 0.55 \cdot 10^{-16} s^{-1}.
\end{equation}

 If the stochastic field is assumed to be coupled to the particle mass density (mass proportional CSL model) (see for example \cite{bassi}) then the previous expression for the emission rate eqn. (\ref{furate}) for electrons needs to be multiplied by the factor $(m_e/m_N)^2$, where $m_N$ is the nucleon mass.

Only the four valence electrons were considered to contribute to the measured X-ray emission, as their binding energy is much lower than the emitted photons energy, so they can be considered as \emph{quasi-free}.

In Ref. \cite{Adler} the author argues that, in evaluating this numerical result, Fu uses for the electron charge the value $e^2=17137.04$, whereas the standard adopted Feynman rules require the identification of $e^2/(4\pi)=17137.04$. We took into account this correction when evaluating the new limit on the collapse rate parameter presented  below.

\section{The analysis of IGEX data}
\label{analysis}
In order to reduce the possible bias  on the $\lambda$ value introduced by the evaluation of the rate at one single energy bin, we  adopted a different strategy with respect to the analysis performed in \cite{fu}. The X-ray emission spectrum measured by the IGEX collaboration \cite{igex1} was fitted in an energy range $\Delta E =$ $4.5\div 48.5$ KeV $\ll m$, compatible with the non-relativistic assumption (for electrons) used in the calculation of the predicted rate (eqn. (\ref{furate})). IGEX is a low-background experiment dedicated to the double beta neutrinoless ($\beta \beta 0 \nu$) decay research based on the use of  low-activity Germanium detectors. The published data \cite{igex2}, used to extract a new upper limit on the collapse rate parameter, refers to an 80 kg day exposure.

The X-ray spectrum was then fitted with the following function:

\begin{equation}\label{fitfunc}
\frac{d\Gamma (E)}{dE} = \frac{\alpha(\lambda)}{E}.
\end{equation}

\noindent
A $\chi^2$ function was obtained in which the expectation value, for the observed number of counts, bin by bin, follows a Gamma distribution, as derived from the Bayes theorem. The $\chi^2$ function was then minimized, from where we extracted a new value for the $\lambda$ parameter, as reported below.

\section{A new limit on $\lambda$ and discussion of the results}\label{fitpar}
\label{fit results}
\begin{figure}
\centering
\includegraphics[height=4in]{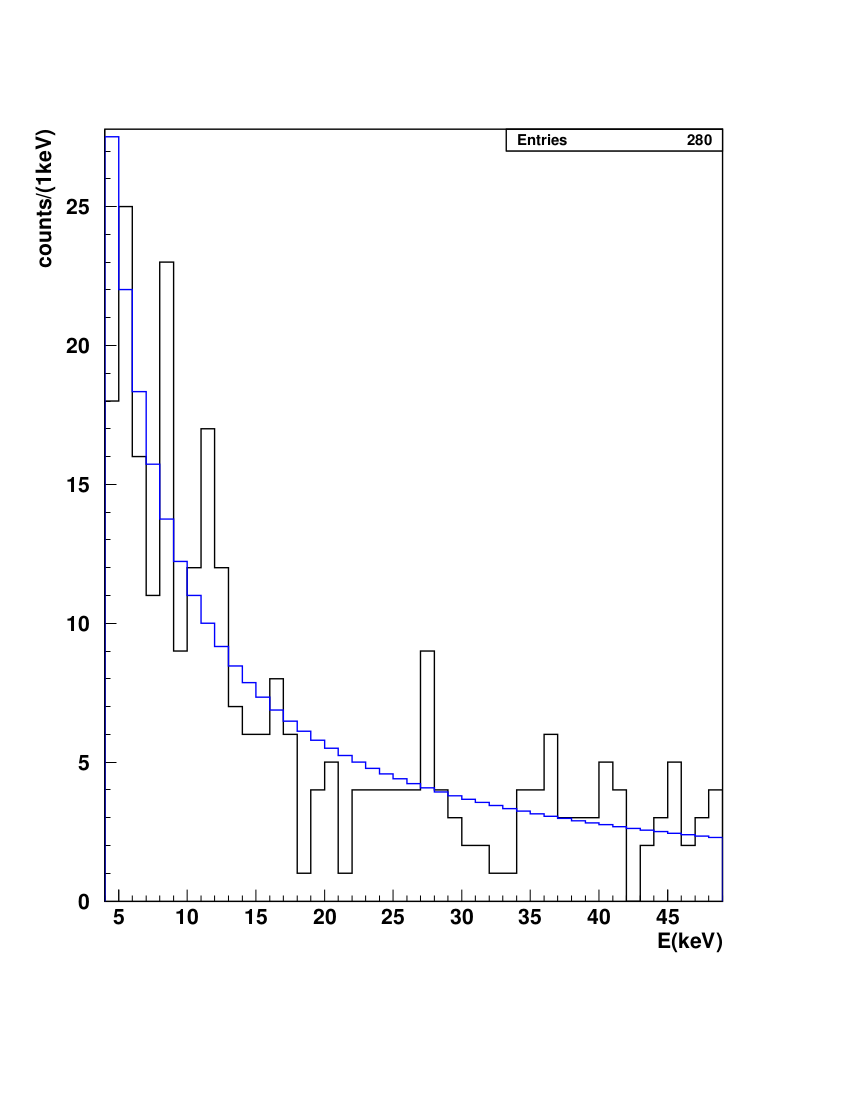}
\caption{Fit of the X-ray emission spectrum measured by the IGEX experiment \cite{igex1,igex2}, using the theoretical fit function given by eqn. (\ref{fitfunc}).}\label{fit}
\end{figure}

The result of the fit performed on the IGEX data is shown in figure \ref{fit}. The minimization gives for the free parameter of the fit the value $\alpha(\lambda) = 110 \pm 7$, corresponding to a reduced chi-square $\chi^2/n.d.f = 1.1$. 

An upper limit on the $\lambda$ parameter can then be set according to eqn. (\ref{furate}):

\begin{equation}\label{fu1}
\frac{d\Gamma (E)}{dE} = c \, \frac{e^2 \lambda}{4\pi^2 a^2 m^2 E}  \leq \frac{110}{E},
\end{equation}
where the $c$ factor is given by:

\begin{equation}
c = (8.9 \,\, 10^{24})\,\,(8.6\,\, 10^4)\,\,(4),
\end{equation}
the first bracket accounts for the particle density of Germanium, the second term is the number of seconds in one day, and the number 4 represents the valence electrons in Germanium. Consistent with Fu's hypothesis, considering the four valence electrons to be quasi-free, applying eqn. (\ref{fu1}) and using the correct prescription $e^2/(4\pi)=17137.04$ (see \cite{Adler}), the following upper limit for the reduction rate parameter was obtained if the stochastic field is assumed to have an universal coupling with all types of particles:

\begin{equation}\label{lim2}
\lambda \leq 1.4 \,\,\, 10^{-17} s^{-1}. 
\end{equation}

If we take into consideration the possible dependence of $\lambda$ on the particle mass, for a mass proportional CSL model, then

\begin{equation}\label{lim3}
 \lambda \leq 4.7 \,\,\, 10^{-11} s^{-1}.
\end{equation}

 The obtained limits improve the previous ones \cite{fu} by a factor 4. These limits  are to be compared with the values originally assumed in \cite{ghi}:

\begin{equation}\label{lmodel}
\lambda_{QMSL} = 10^{-16} s^{-1} \qquad , \qquad   \lambda_{CSL} = 2.2 \,\, 10^{-17} s^{-1}
\end{equation}
and with those proposed, more recently, by S. Adler \cite{Adler}, where $\lambda$ can be as high as $10^{-10}- 10^{-8} s^{-1} $. In the coming years a considerable effort will be dedicated to further improve the limits on the $\lambda$ parameter and  to perform a more stringent test of the collapse models.

\section{Conclusions and  perspectives}
\label{future}
The collapse of the wave function and, more generally, the ``measurement problem'' is  one of the hottest topics in Quantum Mechanics, generating intensive debates and discussions. A possible mechanism inducing the collapse is the so-called continuous spontaneous localisation (CSL), which  has an unique experimental signature: a spontaneous radiation emitted by (free) charged particles. A new limit on the mean collapse frequency parameter  $\lambda$, characterising  the CSL model, was obtained  by performing an analyses of the IGEX experimental data. The $\lambda$ value was obtained to be $\lambda \leq 1.4 \times  10^{-17} s^{-1}$  if no mass dependence is considered, and $\lambda \leq 4.7 \times 10^{-11} s^{-1} $ if, instead, such a dependence is taken into consideration.

We are presently performing a feasibility study to define a dedicated experimental setup which will allow us in the future to improve the limits on the collapse rate parameter $\lambda$ by 2-3 orders of magnitude.
We will then impose very strong constraints on the possible CSL models, being able to exclude those with the $\lambda$-parameter higher than the experimental limit. If, instead, the rate of the emitted radiation, contrary to the expectations, will not go down below a certain limit, this would pose a very interesting problem to the scientific community. 
\begin{acknowledgements}
 We acknowledge the support from the: HadronPhysics FP6(506078), HadronPhysics2 FP7
(227431), HadronPhysics3
(283286) projects, EU COST Action MP1006, Fundamental Problems in
Quantum Physics, Austrian Science Foundation (FWF) which supports the VIP2 project with the
grant P25529-N20 and Centro Fermi (``Problemi aperti nella meccania quantistica'' project).
\end{acknowledgements}



\begin{thebibliography}{}
%
%








\bibitem{ghi} G. C. Ghirardi, A. Rimini and T. Weber, \emph{Phys. Rev. D} \textbf{34} (1986) 47; ibid. 36, 3287 (1987); \emph{Found. Phys.}\textbf{18} (1988) 1. 

\bibitem{pear} P. Pearle,\emph{ Phys. Rev} \textbf{A39} (1989) 2277; G.C. Ghirardi, P. Pearle and A. Rimini \emph{Phys. Rev.} {\bf A42} (1990) 78.


\bibitem{bahrami} M. Bahrami {\it et al.}, \emph{ Sci. Rep.} \textbf{3} (2013) 1952.



\bibitem{fu} Q. Fu,\emph{ Phys. Rev.} \textbf{A56} (1997) 1806.

\bibitem{miley} H. S. Miley, {\it et al.},\emph{ Phys. Rev. Lett.} \textbf{65} (1990) 3092.
3092
\bibitem{igex1} C. E. Aalseth {\it et al.}, IGEX collaboration,\emph{ Phys. Rev. C} \textbf{59}  (1999) 2108.


\bibitem{bassi} A. Bassi, G. C. Ghirardi \emph{Phys. Rep. 379} (2003) 257.


\bibitem{Adler} S. L. Adler, \emph{Journal of Physics A}\textbf{40}  (2007) 2935.


\bibitem{igex2} A. Morales {\it et al.},  IGEX collaboration \emph{Phys. Lett. B}\textbf{532} (2002)  814.






\end{thebibliography}


\end{document}